\documentclass[twocolumn,showpacs,prl,superscriptaddress,preprintnumbers,floatfix]{revtex4}
\usepackage{graphicx}
\newcommand{\ep}{\varepsilon }
\begin{document}
\title{How to reconcile the Rosenbluth and the polarization transfer
methods in the measurement of the proton form factors}
\author{P.A.M. Guichon} 
\affiliation{SPhN/DAPNIA, CEA Saclay, F91191 Gif sur Yvette, France}
\author{M. Vanderhaeghen}
\affiliation{Institut f\"ur Kernphysik, Johannes Gutenberg Universit\"at,
  D-55099 Mainz, Germany}
%
%
\begin{abstract}
The apparent discrepancy between the Rosenbluth and the polarization
transfer method for the ratio of the electric to magnetic proton form
factors can be explained by a two-photon exchange correction which
does not destroy the linearity of the Rosenbluth plot. Though intrinsically
small, of the order of a few percent of the cross section, this correction
is accidentally amplified in the case of the Rosenbluth method. 
\end{abstract}
\maketitle
\begin{figure}
\includegraphics[width=8cm,height=5.5cm]{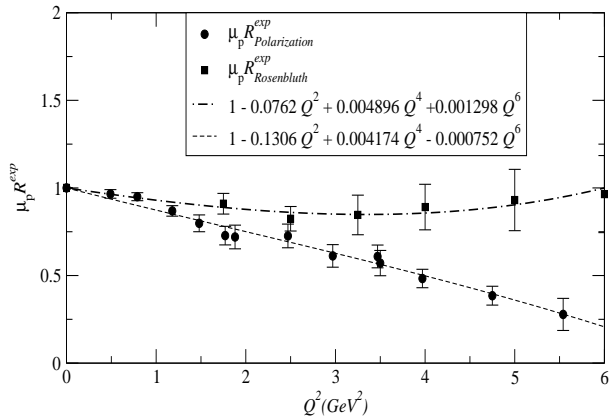}
\caption{Experimental values of 
\protect\(R^{exp}_{Rosenbluth}\protect \)~\cite{LT1994}
and \protect\( R^{exp}_{Polarization}\protect \)~\cite{Jones00,Gayou02} 
and their polynomial fits.}
\label{Fig:fit}
\end{figure}

\begin{figure}[h]
\includegraphics[width=5.5cm]{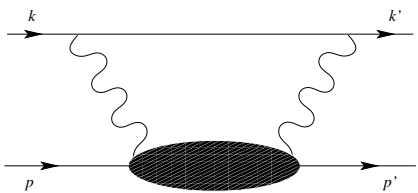}
\caption{The box diagram. The filled blob represents the response
of the nucleon to the scattering of the virtual photon.}
\label{Fig:box}
\end{figure}

 
%
%
The electro-magnetic form factors are essential pieces of our knowlegde
of the nucleon structure and this justifies the efforts devoted to
their experimental determination. They are defined as the matrix elements
of the electro-magnetic current \( J^{\mu }(x) \) according to 
~:
\begin{eqnarray}
\label{Eq:intro.1}
&&<N(p')|J^{\mu }(0)|N(p)>  \nonumber\\
&&= \, e\, \bar{u}(p')\left[ G_{M}(Q^{2})\gamma ^{\mu }-F_{2}(Q^{2})\frac{(p+p')^\mu}{2M}\right] u(p),
\end{eqnarray}
where \( e\simeq \sqrt{4\pi /137} \) is the proton charge, 
\( M \) the nucleon mass, 
and $Q^2$ the squared momentum transfer.  
The magnetic form factor \( G_{M} \) is related
to the Dirac (\( F_{1}) \) and Pauli (\( F_{2}) \) form 
factors by \( G_{M}=F_{1}+F_{2} \), 
and the electric form factor is given by \( G_{E}=F_{1}-\tau F_{2}
\), with \( \tau =Q^{2}/4M^{2} \).  
For the proton, $F_{1}(0)=1$, and $F_{2}(0)=\mu _{p}-1=1.79$.
In the one-photon exchange or Born approximation, 
elastic lepton-nucleon scattering~:
\begin{equation}
\label{Eq:intro.2}
l(k)+N(p)\rightarrow l(k')+N(p'),
\end{equation}
gives direct access to the form factors in the spacelike region 
($ Q^{2}>0)$, through its cross section~: 
\begin{equation}
\label{Eq:intro.5}
d\sigma _{B}=C_{B}(Q^{2},\varepsilon )
\left[ G_{M}^{2}(Q^{2})+\frac{\varepsilon }{\tau }G_{E}^{2}(Q^{2})\right] ,
\end{equation}
where $\varepsilon$ is the photon polarization parameter, 
and \( C_{B}(Q^{2},\varepsilon ) \) is a phase space factor which is 
irrelevant in what follows. 
For a given value of \( Q^{2} \), Eq.~(\ref{Eq:intro.5}) shows that it is
sufficient to measure the cross section for two values of \( \varepsilon  \)
to determine the form factors \( G_{M} \) and \( G_{E}\). 
This is referred to as the Rosenbluth method~\cite{Rosenb50}.
The fact that \( d\sigma /C_{B}(Q^{2},\varepsilon ) \) is a linear
function of \( \varepsilon  \) (Rosenbluth plot criterion) is generally
considered as a test of the validity of the Born approximation. 
\newline
\indent
Polarized lepton beams give another way to access the 
form factors~\cite{Akhieser58}.
In the Born approximation, the polarization of the recoiling proton
along its motion (\( P_{l} \)) is proportional to \( G_{M}^2 \) while
the component perpendicular to the motion (\( P_{t} \) ) is proportional
to \( G_E G_M\). We call this the polarization method for short. Because
it is much easier to measure ratios of polarizations, it has been
used mainly to determine the ratio \( G_{E}/G_{M} \) through a measurement
of \( P_{t}/P_{l} \) using~\cite{Akhiezer68}~:
\begin{equation}
\label{Eq:intro.8}
\frac{P_{t}}{P_{l}}=-\sqrt{\frac{2\varepsilon }{\tau (1+\varepsilon )}}\frac{G_{E}}{G_{M}}.
\end{equation}
\indent
Thus, in the framework of the Born approximation, one has two independent
measurements of \( R=G_{E}/G_{M}. \) In Fig. \ref{Fig:fit} we show
the corresponding results, 
which we call \( R^{exp}_{Rosenbluth} \)
and \( R^{exp}_{Polarization} \) , for the range of \( Q^{2} \)
which is common to both methods. 
The data are from Refs.~\cite{LT1994,Jones00,Gayou02}.
It is seen that the deviation between the two methods starts 
around $Q^2 = 2$~GeV$^{2}$ and increases with \( Q^{2} \), 
reaching a factor \( 4 \) at  $Q^2 = 6$~GeV$^{2}$.
A recent re-analysis of the SLAC cross sections~\cite{Arr03} 
and new  Rosenbluth measurements from JLab~\cite{Chr03} 
confirm that the Rosenbluth and polarization extractions
of the ratio $ G_{E}/G_{M} $ are incompatible at large $Q^2$.
This discrepancy is a serious problem as it generates confusion and
doubt about the whole methodology of lepton scattering experiments.
\newline
\indent
In this letter we take a first step to unravel this problem by interpreting
the discrepancy as a failure of the Born approximation which nevertheless
does not destroy the linearity of the Rosenbluth plot. This means
that we give up the beloved one-photon exchange concept and enter
the not well paved path of multi-photon physics. By this we do not
mean the effect of soft (real or virtual) photons, that is the radiative
corrections. The effect of the latter is well under control because
their dominant (infra-red) part can be factorized in the observables
and therefore does not affect the ratio \( G_{E}/G_{M} \). Here we
consider genuine exchange of hard photons between the lepton and the
hadron. Such higher-order corrections to the one-photon exchange
approximation have been considered in the past \cite{Dr59,Gr69}, 
and their effects were found to be of order 1 - 2 \% on the cross section. 
However, such estimates based on nucleon and resonance
intermediate states can only be expected to give a realistic description
of the nucleon structure for momentum transfers up to $Q^2 \lesssim
1$~GeV$^2$, whereas they are largely unknown at higher values of $Q^2$.
\newline
\indent
Even if we restrict ourselves to the two-photon exchange case, the evaluation
of the box diagram (Fig.~\ref{Fig:box}) involves the \emph{full} reponse
of the nucleon to doubly virtual Compton scattering and we do not
know how to perform this calculation in a model independent way. Therefore
we adopt a modest strategy based on the phenomenological consequences
of using the full \( eN \) scattering amplitude rather than its Born
approximation. Though it cannot lead to a full answer it produces
the following interesting results:
\begin{itemize}
\item the two-photon exchange amplitude needed to explain the discrepancy
is actually of the expected order of magnitude, that is a few percent
of the Born amplitude.
\item there may be a simple explanation of the fact that the Rosenbluth
plot looks linear even though it is strongly affected by the two-photon
exchange.
\item the polarization method result is little affected by the two-photon
exchange, \emph{at least in the range of \( Q^{2} \) which has been
studied until now.}
\end{itemize}

\begin{figure}
\includegraphics[width=8.25cm,height=5.75cm]{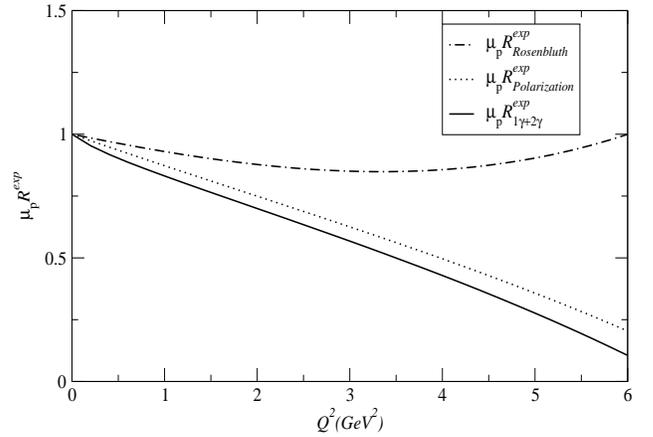}
\caption{Comparison of the experimental ratios \protect\( \mu _{p}R^{exp}_{Rosenbluth}\protect \)
and \protect\( \mu _{p}R^{exp}_{Polarization}\protect \) with the
value of \protect\( \mu _{p}R^{exp}_{1\gamma +2\gamma } \protect \)
deduced from our analysis.}
\label{X_vs_exp}
\end{figure}

To proceed with the general analysis of elastic electron-nucleon
scattering (\ref{Eq:intro.2}), we adopt the usual definitions~:
\begin{equation}
\label{Eq:intro.3}
P=\frac{p+p'}{2},\, K=\frac{k+k'}{2},\, q=k-k'=p'-p,
\end{equation}
and choose \begin{equation}
\label{Eq:intro.4}
Q^{2}=-q^{2},\, \nu =K.P ,
\end{equation}
as the independent invariants of the scattering. 
The polarization parameter $\varepsilon$ of the virtual photon 
is related to the invariant $\nu$ as (neglecting the electron mass 
$m_e$)~:
\begin{equation}
\label{Eq:intro.6}
\varepsilon =\frac{\nu ^{2}-M^{4}\tau (1+\tau )}{\nu ^{2}+M^{4}\tau (1+\tau )}.
\end{equation}
\indent
For a theory which respects Lorentz, parity and charge
conjugation invariance, the $T$-matrix for elastic scattering
of two spin 1/2 particles can be expanded
in terms of six independent Lorentz structures which, 
following Ref.~\cite{Goldb57}, can be chosen as:
\(\bar{u}(k')u(k)\ \bar{u}(p')u(p)\),
\(\bar{u}(k')u(k)\ \bar{u}(p')\gamma .Ku(p)\),
\(\bar{u}(k')\gamma _{5}u(k)\ \bar{u}(p')\gamma _{5}u(p)\), 
\(\bar{u}(k')\gamma .Pu(k)\ \bar{u}(p')\gamma .Ku(p)\), 
\(\bar{u}(k')\gamma .Pu(k)\ \bar{u}(p')u(p)\),
\(\bar{u}(k')\gamma _{5}\gamma .Pu(k)\ \bar{u}(p')\gamma _{5}\gamma . Ku(p)\). 
In the limit \(m_e\to 0\),  the vector nature of
the coupling in QED implies that any Feynman diagram is invariant 
under the chirality operation
\(u(k)\to\gamma_5 u(k), \bar{u}(k')\to -\bar{u}(k')\gamma_5 \). 
Therefore the Lorentz structures which change their sign 
under this operation must come with an explicit factor \(m_e\). 
This allows us to neglect the structures which contain either 
\(\bar{u}(k')u(k)\) or \(\bar{u}(k')\gamma_5 u(k)\). 
Using the Dirac equation and elementary relations 
between Dirac matrices the linear combination of the remaining 
three amplitudes can be written in the form~:
\begin{eqnarray}
\label{Eq:Adec.2}
T \,&=&\, \frac{e^{2}}{Q^{2}} \, \bar{u}(k')\gamma _{\mu }u(k)\, 
\nonumber\\
&\times& \bar{u}(p')\left( \tilde{G}_{M}\, \gamma ^{\mu }
-\tilde{F}_{2}\frac{P^{\mu }}{M}
+\tilde{F}_{3}\frac{\gamma .KP^{\mu }}{M^{2}}\right) u(p),
\end{eqnarray}
where \( \tilde{G}_{M},\, \tilde{F}_{2},\, \tilde{F}_{3} \) are 
complex functions of \( \nu  \) and \( Q^{2} \), and 
where the factor \( e^{2}/Q^{2} \)
has been introduced for convenience. 
In the Born approximation, one obtains~:
\begin{eqnarray}
\label{Eq:Adec.4}
\tilde{G}_{M}^{Born}(\nu ,Q^{2}) \,&=&\, G_{M}(Q^{2}),   \nonumber\\
\tilde{F}_{2}^{Born}(\nu ,Q^{2}) \,&=&\, F_{2}(Q^{2}),   \nonumber\\
\tilde{F}_{3}^{Born}(\nu ,Q^{2}) \,&=&\, 0. 
\end{eqnarray}
Since \( \tilde{F}_{3} \) and the phases of \( \tilde{G}_{M} \)
and \( \tilde{F}_{2} \) vanish in the Born approximation, they must
originate from processes involving at least the exchange of two photons.
Relative to the factor \( e^{2} \) introduced in 
Eq.~(\ref{Eq:Adec.2}), we see that they are at least of order \( e^{2}. \)
This, of course, assumes that the phases of \( \tilde{G}_{M} \) and
\( \tilde{F}_{2} \) are defined, which amounts to supposing that, in
the kinematical region of interest, the moduli of \( \tilde{G}_{M} \)
and \( \tilde{F}_{2} \) do not vanish, which we take for granted in
the following.  
Defining~:  
\begin{equation}
\label{Eq:Obs.1}
\tilde{G}_{M}=e^{i\phi _{M}} | \tilde{G}_{M} | ,\, 
\tilde{F}_{2}=e^{i\phi _{2}} | \tilde{F}_{2} | ,\, 
\tilde{F}_{3}=e^{i\phi _{3}} | \tilde{F}_{3} | ,
\end{equation}
and using standard techniques, we get the following expressions
for the observables of interest~:
\begin{eqnarray}
d\sigma &=& C_{B}(\nu ,Q^{2}) \, \frac{\ep (1+\tau )}{\tau }  \nonumber\\
&\times& \left\{ | \tilde{G}_{M}| ^{2} \,
\frac{\rho ^{2}-\tau +\tau ^{2}}{\rho ^{2}-\tau -\tau ^{2}}
+ | \tilde{F}_{2}| ^{2}(1+\tau )\right. \nonumber \\
&& -2 \, | \tilde{G}_{M} | \, \left( \cos \phi _{2M} 
| \tilde{F}_{2}| -\cos \phi _{3M} | \tilde{F}_{3}| \rho \right) \nonumber\\
&& \left. -2 \, \cos \phi _{23} \, | \tilde{F}_{2}\tilde{F}_{3}| \rho
+ | \tilde{F}_{3} | ^{2}(\rho ^{2}-\tau ^{2}) \right\} ,
\label{Eq:Obs.2} \\
\frac{P_{t}}{P_{l}} & =& 
-\sqrt{\frac{\rho ^{2}-\tau -\tau ^{2}}{\tau }} \nonumber\\
\times && \hspace{-0.5cm}
\frac{ | \tilde{G}_{M} | 
-\cos \phi _{2M} | \tilde{F}_{2} | (1+\tau )
+\cos \phi _{3M} | \tilde{F}_{3} | \rho }
{ | \tilde{G}_{M} | \rho 
+\cos \phi _{3M} | \tilde{F}_{3}| (\rho ^{2}-\tau -\tau^{2})},
\label{Eq:Obs.3} 
\end{eqnarray}
with 
$\phi _{2M}=\phi _{2}-\phi _{M}$, 
$\phi _{3M}=\phi _{3}-\phi _{M}$, 
$\phi _{23}=\phi _{2}-\phi _{3}$, and $\rho = \nu / M^2$.
If one substitutes the Born approximation values of the amplitudes
(\ref{Eq:Adec.4}) then Eqs.~(\ref{Eq:Obs.2},\ref{Eq:Obs.3}) give
back the familiar expressions of Eqs.~(\ref{Eq:intro.5},\ref{Eq:intro.8}). 
\newline
\indent
To simplify the above general expressions,
we make the very reasonable assumption that only the two-photon exchange
needs to be considered. In practice we make an expansion in power of \(e^2\) 
 of Eqs.~(\ref{Eq:Obs.2},\ref{Eq:Obs.3}) using the fact that 
 \( \phi _{M},\, \phi _{2} \) and \( \tilde{F}_{3} \)
are at least of order \( e^{2} \) but we do not expand \(| \tilde{G}_{M}|\) and \(| \tilde{F}_{2}|\), which 
 is perfectly legitimate. 
This leads to  the following approximate expressions~:
\begin{eqnarray}
\frac{d\sigma }{C_{B}(\varepsilon ,Q^{2})}
&\simeq&\frac{| \tilde{G}_{M} | ^{2}}{\tau } 
\left\{  \tau +\varepsilon  
\frac{|\tilde{G}_{E}|^2}{|\tilde{G}_{M}|^2} \right. \nonumber\\
&&\hspace{-0.5cm} \left. + \,2 \varepsilon \left( \tau +
\frac{|\tilde{G}_{E}|}{|\tilde{G}_{M}|} \right) 
{\cal R}\left( \frac{\nu \tilde{F}_{3}}{M^{2}|\tilde{G}_{M}| }\right)\right\}, 
\label{Eq:Obs.9} 
\end{eqnarray}
\begin{eqnarray}
\frac{P_{t}}{P_{l}} & \simeq & - \, 
\sqrt{\frac{2\varepsilon }{\tau(1+\varepsilon )}} \,
\left\{\, \frac{|\tilde{G}_{E}|}{|\tilde{G}_{M}|} \right. \nonumber\\
&&\left. +\left( 1-\frac{2\varepsilon }{1+\varepsilon }
\frac{|\tilde{G}_{E}|}{|\tilde{G}_{M}|} \right) {\cal
R}\left( \frac{\nu \tilde{F}_{3}}{M^{2} | \tilde{G}_{M} |}
\right) \right\} ,
\label{Eq:Obs.10} 
\end{eqnarray}
where the neglected terms are of order $e^4$ w.r.t. the leading one. 
By analogy, we have defined~:
\begin{equation}
\label{Eq:Obs.11}
\tilde{G}_{E}=\tilde{G}_{M}-(1+\tau )\tilde{F}_{2},\, 
\end{equation}
and \( {\cal R} \) denotes the real part. 
Note that $\tilde{G}^{Born}_{E}(\nu, Q^2)=G_{E}(Q^2)$. 
To set the scale for the size of the two-photon exchange term 
\( (\tilde{F}_{3}) \)
we introduce the dimensionless ratio~:
\begin{equation}
Y_{2\gamma }(\nu, Q^{2})={\cal R}\left( \frac{\nu
\tilde{F}_{3}}{M^{2} | \tilde{G}_{M} | }\right) .
\end{equation}
In the region of large \(Q^2\) which is where 
the discrepancy really gets large, \(\tau\) is of
order 1 or larger, 
while we can take as upper bound estimate 
\( | \tilde{G}_{E} | / | \tilde{G}_{M} |\simeq
G_E(0)/G_M(0)=1/2.79   \) . So,  for a qualitative reasoning, we can neglect 
\( | \tilde{G}_{E} | / | \tilde{G}_{M}| \) with respect to \(\tau\) 
and, up to a term quadratic in
\(Y_{2\gamma }\), the cross section has the form 
\(| \tilde{G}_{M} | ^{2}(1+ \varepsilon \, Y_{2\gamma })^{2} \). 
So we expect \(Y_{2\gamma }\sim \alpha \simeq 1/137\). 
However in the Rosenbluth method where 
one identifies \((G_E/G_M)^2\) with the coefficient of \( \varepsilon\),  
the two photon effect comes as a correction 
to a small number \(\sim (1/2.79)^2 \). 
So we expect  that the correction will have a stronger effect 
in the Rosenbluth than in the polarization method.
\newline
\indent
From Eqs.~(\ref{Eq:Obs.9},\ref{Eq:Obs.10}) we see that the 
pair of observables \( (d\sigma ,\, P_{t}/P_{l}) \) 
depends on $| \tilde{G}_{M}|$, 
$| \tilde{G}_{E}|$, and $ {\cal R}(\tilde{F}_{3})$. 
In the first approximation, we know
that $| \tilde{G}_{M}(\nu , Q^{2})| \simeq G_{M}(Q^{2})$, 
$| \tilde{G}_{E}(\nu , Q^{2})| \simeq G_{E}(Q^{2})$, 
and only \( {\cal R}(\tilde{F}_{3}) \) is really a new unknown parameter.
Thus allowing for two-photon exchange somewhat complicates the interpretation
of the lepton scattering experiments but not in a dramatic way. 
The main uncertainty is the dependence on \(\nu\) 
(or \(\ep\)) of
\(\tilde{F}_3\) and to further simplify the problem we make the
following observations. First, if we look at the data of 
Ref.~\cite{LT1994} for \( d\sigma /C_{B}(\varepsilon ,Q^{2}) \)\
as a function of \( \varepsilon  \)
we observe that for each value of \( Q^{2} \) the set of points are
pretty well aligned. We see in Eq. (\ref{Eq:Obs.9}) that this can
be understood if, at least in the first approximation, 
the product \( \nu \, \tilde{F}_{3} \)
is independent of \( \varepsilon . \) We do not have a first principle
explanation for this but we feel allowed to take it as experimental
evidence. To explain the linearity of the plot one must also suppose
that \( | \tilde{G}_{M}|  \) and \( | \tilde{G}_{E}|  \)
are independent of \( \ep  \) (that is \( \nu  \) ) but since the
dominant term of these amplitudes depends only on \( Q^{2} \) this
is a very mild assumption. We then see from Eq.~(\ref{Eq:Obs.9}) that
what is measured using the Rosenbluth method is~:
\begin{equation}
\label{Eq:Ana.1}
(R^{exp}_{Rosenbluth})^{2}=\frac{|\tilde{G}_{E}|^{2}}{|\tilde{G}_{M}|^{2}}
+2\left( \tau +\frac{| \tilde{G}_{E}| }{| \tilde{G}_{M}| }\right) Y_{2\gamma },
\end{equation}
with \( | \tilde{G}_{E}| / | \tilde{G}_{M}|  \)
and \( Y_{2\gamma } \) essentially independent of \( \ep  \) , rather
than 
$(R^{exp}_{Rosenbluth})^{2}= ( G_{E} / G_{M} ) ^{2}$~,
as implied by one-photon exchange. 
Second, the experimental results of the polarization method
have been obtained for a rather narrow range of 
\( \varepsilon  \), typically 
from \( \varepsilon = 0.75 \) to \( 0.9 \) 
for the points at large \(Q^2\).
So, in practice, we can neglect
the \( \ep  \) dependence of \( R^{exp}_{Polarization} \) and from
Eq.~(\ref{Eq:Obs.10}) we see that this experimental ratio must be
interpreted as~: 
\begin{equation}
\label{Eq:Ana.3}
R^{exp}_{Polarization}=\frac{|\tilde{G}_{E}|}{|\tilde{G}_{M}|} 
+\left( 1-\frac{2\varepsilon }{1+\varepsilon } 
\frac{|\tilde{G}_{E}|}{|\tilde{G}_{M}|} \right) Y_{2\gamma },
\end{equation}
rather than $R^{exp}_{Polarization}= G_{E} / G_{M}$.
In order that Eq.~(\ref{Eq:Ana.3}) be consistent with our hypothesis
we should find that \( Y_{2\gamma } \) is small enough that the factor
\( 2\ep /(1+\ep ) \) introduces no noticeable \( \ep  \) dependence
in \( R^{exp}_{Polarization} \).
\begin{figure}
\includegraphics[width=8cm,height=5.6cm]{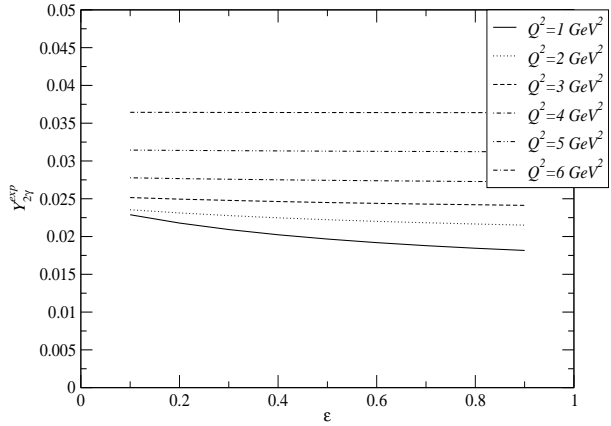}
\caption{The ratio \protect\( Y^{exp}_{2\gamma }\protect \) versus 
\protect\( \varepsilon \protect \) 
for several values of \protect\( Q^{2}\protect \).}
\label{ratioY}
\end{figure}
\newline
\indent
We can now solve Eqs.~(\ref{Eq:Ana.1},\ref{Eq:Ana.3})
for \( | \tilde{G}_{E}| / | \tilde{G}_{M}|  \)
and \( Y_{2\gamma } \) for each \( Q^{2} \) .
Since the system of equations is equivalent to a quadratic
equation it is more efficient to solve it numerically. For this we
have fitted the data by a polynomial in \( Q^{2} \) as
shown in Fig.~\ref{Fig:fit}, and we shall consider this fit as the
experimental values. In particular we do not attempt to represent
the effect of the error bars which can be postponed to a more complete
re-analysis of the data.
The solution of Eqs.~(\ref{Eq:Ana.1},\ref{Eq:Ana.3})
for the ratio \( Y^{exp}_{2\gamma } \) is shown in Fig.
\ref{ratioY} where we can see that, 
as expected, it is essentially flat as a function of \(\varepsilon\) 
and small,  of the order of a few percent. 
Thus a tiny correction allows the Rosenbluth and the
polarization method to give the same value for 
\(| \tilde{G}_{E}| / | \tilde{G}_{M}|\). It is reasonable to 
think that \( \delta G_E=\tilde{G}_{E}-G_{E} \) 
and \( \delta G_M=\tilde{G}_{M}-G_{M}\) are 
comparable to  \( Y^{exp}_{2\gamma } \)  
and therefore \(| \tilde{G}_{E}| / | \tilde{G}_{M}|\) should not 
be very different from the actual value of \(G_E/G_M\).
So it makes sense to compare the value we get for 
\( R^{exp}_{1\gamma +2\gamma } = | \tilde{G}_{E}| / | \tilde{G}_{M}|\)
with the starting experimental ratios \( R^{exp}_{Rosenbluth} \)
and \( R^{exp}_{Polarization} \). This is shown 
in Fig. \ref{X_vs_exp}, from which we see that
\( R^{exp}_{1\gamma +2\gamma } \) is close to \( R^{exp}_{Polarization} \). 
The difference between the two curves can be attributed  either to  \(
Y^{exp}_{2\gamma } \) or to
\( (\delta G_M,\ \delta G_E) \). Insofar as 
\( (\delta G_M,\ \delta G_E) \) are of the same order of magnitude as
 \( Y^{exp}_{2\gamma } \), which is small according to our analysis, our
interpretation of this small difference is that the polarization 
method is little affected by the two-photon correction.
\newline
\indent
In summary, the discrepancy between the Rosenbluth and the polarization
method for \( G_{E}/G_{M} \) can be attributed to a failure of the
one-photon approximation which is amplified at large
\( Q^{2} \) in the case of the Rosenbluth method.  
The expression for the cross section 
also suggests that the two-photon effect does not destroy the linearity
of the Rosenbluth plot provided the product \( {\cal R}(\nu \tilde{F}_{3}) \)
is independent of \( \nu . \) It remains to be investigated if there
is a fundamental reason for this behavior or if it is fortuitous. 
Using the existing data we have extracted the essential piece of the
puzzle, that is the ratio \( Y^{exp}_{2\gamma } \) which measures
the relative size of the two-photon amplitude \( \tilde{F}_{3} \).
Within our approximation scheme, we find that \( Y^{exp}_{2\gamma } \)
is of the order of a few percent. 
This is a very reassuring result since 
this is the order of magnitude expected for two-photon corrections.
What is needed next  
is a realistic evaluation of this particular amplitude. A first step
in this direction was performed very recently in Ref.~\cite{Blund03},
where the contribution to the two-photon exchange amplitude 
was calculated for a nucleon intermediate state in
Fig.~\ref{Fig:box}. 
The calculation of Ref.~\cite{Blund03} 
found that the two-photon exchange correction with
intermediate nucleon has the proper sign and magnitude to resolve a
large part of the discrepancy between the two experimental techniques,
confirming the finding of our general analysis. As a next step, an
estimate of the inelastic part is needed to fully quantify the 
nucleon response in the two-photon exchange process. 
\newline
\indent 
From our analysis we extract 
the ratio \( | \tilde{G}_{E}| / | \tilde{G}_{M}|  \)
which in the first approximation should not be very different from 
\( G_{E}/G_{M} \).
We find that it is close to the value obtained by the polarization
method when one assumes the one-photon exchange approximation. 
This comparison is meaningful if, 
as suggested by the smallness of \( Y^{exp}_{2\gamma } \), 
\(\delta G_E\) and \(\delta G_M\) are negligible.
This could be checked 
by a realistic calculation of the two-photon corrections.  
However we think that a definitive conclusion will wait for the 
determination of \(\delta G_E\) and \(\delta G_M\) as we did for 
 \( Y^{exp}_{2\gamma } \). The necessary experiments probably require 
the use of positrons as well as electron beams. 
\newline
\indent
This work was supported 
by the French Commissariat \`a l'Energie Atomique (CEA),  
and by the Deutsche Forschungsgemeinschaft (SFB443).

\end{document}